\documentclass[12pt,thmsa]{article}
\usepackage{amssymb}

\usepackage{sw20lart}

\input{tcilatex}
\input tcilatex
\begin{document}

\author{Emilio Santos \\
Departamento de F\'{i}sica \and Universidad de Cantabria. Santander. Spain.
\and FAX: 34-42-201402. email: santose@unican.es}
\title{Space-time curvature due to quantum vacuum fluctuations, an alternative to
dark energy ?}
\date{October, 7, 2009}
\maketitle

\begin{abstract}
It is pointed out that quantum vacuum fluctuations may give rise to a
curvature of space-time equivalent to the curvature currently attributed to
dark energy. A simple calculation is made, involving plausible assumptions
within the framework of quantized gravity, which suggests that the value of
the dark energy density is roughly given by the product of Newton\'{}s
constant times the quantity m$^{6}$c$^{4}$
\rlap{\protect\rule[1.1ex]{.325em}{.1ex}}h%
$^{-4},$ m being a typical mass of elementary particles. The estimate is
compatible with observations.


\end{abstract}

The observed accelerated expansion of the universe\cite{Sahni} is assumed to
be due to a positive mass density and negative pressure constant throughout
space and time, which is popularly known as ``dark energy''. The mass
density, $\rho _{DE},$ and the presure, $p_{DE},$ associated with the dark
energy are \cite{WMAP} 
\begin{equation}
\rho _{DE}\simeq -p_{DE}\simeq 10^{-26}\text{ kg/m}^{3}.  \label{1}
\end{equation}
The current wisdom is to identify the dark energy with the cosmological
constant introduced by Einstein in 1917 or, what is equivalent in practice,
to assume that it corresponds to the quantum vacuum. Indeed the equality $%
\rho _{DE}=-p_{DE}$ is appropriate for the vacuum (in Minkowski space, or
when the space-time curvature is small) because it is invariant under
Lorentz transformations.

A problem appears when one attempts to estimate the value of $\rho _{DE}$%
\cite{W}. In fact if the dark energy is really due to the quantum vacuum it
seems difficult to understand why the mass density is not either strictly
zero or of order Planck\'{}s density, that is 
\begin{equation}
\rho _{DE}\approx \frac{c^{5}}{G^{2}
\rlap{\protect\rule[1.1ex]{.325em}{.1ex}}h%
}\simeq 10^{97}\text{ kg/m}^{3},  \label{2}
\end{equation}
which is about 123 orders of magnitude larger than eq.$\left( \ref{1}\right) 
$. On the other hand it is known that the correct order is obtained using
the following combination of the fundamental constants $
\rlap{\protect\rule[1.1ex]{.325em}{.1ex}}h%
$ and $c$ with some mass, $m$, typical of elementary particles\cite{Z} 
\begin{equation}
\rho _{DE}\approx G\frac{m^{6}c^{2}}{
\rlap{\protect\rule[1.1ex]{.325em}{.1ex}}h%
^{4}}.  \label{40}
\end{equation}
The observed value, eq.$\left( \ref{1}\right) ,$ is obtained if the mass $m$
is 
\[
m\sim 7.6\times 10^{-29}kg 
\]
which is about 1/20 times the proton mass or about 80 times the electron
mass. I believe that the agreement between eqs.$\left( \ref{1}\right) $ and $%
\left( \ref{40}\right) $ is not an accident and the purpose of this letter
is to propose a possible explanation. More specifically the aim will be to
explain why the density of dark energy may be obtained as a product of
Newton constant, $G$, times some expression involving fundamental parameters
but not $G$. If this is the case, dimensional considerations lead to eq.$%
\left( \ref{40}\right) $ or an equivalent expression with some
characteristic length or time, instead of a mass, in addition to the
constants $
\rlap{\protect\rule[1.1ex]{.325em}{.1ex}}h%
$ and $c$.

We might rewrite eq.$\left( \ref{40}\right) $ in the form

\begin{equation}
\rho _{DE}\approx G\frac{m^{2}}{l}\times \frac{1}{l^{3}},l=\frac{%
\rlap{\protect\rule[1.1ex]{.325em}{.1ex}}h%
}{mc},  \label{41}
\end{equation}
which suggests looking at the dark energy as a gravitational energy per unit
volume\cite{Z}. Here I propose a different interpretation which follows from
looking, not directly at the dark energy itself, but at the curvature of
space-time attributed to the dark energy. Indeed what is actually derived
from astronomical observations is the curvature of space-time\cite{Sahni}.
It is associated to an Einstein tensor, $G_{\mu }^{\nu },$ with components
(from now on I shall use units such that $c=1$). $\;$

\begin{equation}
G_{4}^{4}=G_{1}^{1}=G_{2}^{2}=G_{3}^{3}=8\pi G\,\rho _{DE},\;  \label{42}
\end{equation}
all non-diagonal elements of $G_{\mu }^{\nu }$ being zero. (In eq.$\left( 
\ref{42}\right) $ I have ignored the contribution of matter - either
baryonic or dark- and radiation.) If we take seriously eq.$\left( \ref{40}%
\right) ,$ its combination with eq.$\left( \ref{42}\right) $ tells us that%
\textit{\ }the Einstein tensor, currently\ attributed\ to\ dark\ energy,\
is\ the product of\ the\ square\ of\ Newton$\acute{}$s\ constant,\ $G,$
times a tensor which would depend on properties of quantum fields excluding
gravity. (Of course the statement does not apply to the early universe.) In
practice the Einstein tensor is treated as classical and derivable from a
metric tensor also classical. Thus, if eq.$\left( \ref{40}\right) $ is
meaningfull, \textit{dark energy is associated to a metric tensor which
departs from Minkowski\'{}s by terms of order G}$^{2}$. This is the fact
that I will attempt to explain in the present letter.

My essential assumption is that dark energy is a consequence of the quantum
vacuum fluctuations. The assumption has been considered previously\cite{pad}%
, but the treatment here is different. In order to explain why vacuum
fluctuations lead to a metric\ tensor\ which departs\ from\ Minkowski\'{}s\
by\ terms\ of\ order\ G$^{2}$, I shall start recalling a well known
prediction of quantum mechanics, namely that correlations between quantum
fluctuations may produce observable effects of second order in the coupling
constant. An illustrative example is the (van der Waals) interaction between
two molecules at a distance, $d$, much bigger than the typical size of a
molecule. If both molecules possess a permanent electric dipole moment, then
at low enough temperature they are oriented so that the molecules attract
each other. In fact there is a dipole-dipole (negative) interaction energy
which scales as $d^{-3}.$ Now let us consider two neutral molecules which do
not possess permanent dipole moment. In this case classical physics predicts
that there is no electrostatic interaction between them. Quantum theory
however predicts an interaction due to the fact that quantum fluctuations
give rise to instantaneous dipole moments in both molecules, which are
correlated so that the energy is a minimum. This leads to an interaction
which scales precisely as the square of the coupling parameter above
mentioned, that is $\left( d^{-3}\right) ^{2}=d^{-6}.$ Indeed it is well
known that the interaction energy between nonpolar molecules decreases with
the six power of the distance (when retardation effects are negligible). The
general behaviour may be understood via perturbation theory. To first order
an average of the quantum fluctuations appears, which is zero. To second
order however the perturbation involves the product of two correlated
fluctuations, which is not zero and it gives the interaction to lowest
order. The square of the quantum fluctuations involves the coupling constant
squared. I propose that a similar phenomenon should appear in gravity.

More explicitly stated, the basic hypothesis will be that matter may be
described by a set of interacting quantum fields and that an energy-momentum
quantum tensor operator, $\widehat{T}_{\mu }^{\nu },$ associated to these
fields makes sense. Furthermore I will assume that the expectation of the
said tensor operator, in the quantum state of the universe, $\mid \Phi >,$
may be written as a sum of two contributions. The first one derives from
baryonic matter, dark matter and radiation. That expectation may be treated
as a classical energy-momentum tensor (except in the early universe). The
second contribution will be the vacuum expectation value of the operator, $%
\widehat{T}_{\mu }^{\nu }.$ In order to simplify the treatment that follows
I will ignore the former contribution (including it is straightforward, but
the arguments would be more involved). With respect to the latter, I will
assume that its ``true'' value is zero or negligible in cosmology, although
the possibility that this is not the case will be briefly discussed at the
end of the present letter. (Here I use the word ``true'' in the same sense
as Zel\'{}dovich\cite{Z2}. That is the true part of the vacuum energy is
what remains after subtracting effectively some contribution by means of a
renormalization of Newton\'{}s constant.) That is, for the moment I shall
assume that, at any space-time point, 
\begin{equation}
\left\langle 0\left| \widehat{T}_{\mu }^{\nu }\left(
x^{1},x^{2},x^{3},x^{4}\right) \right| 0\right\rangle =0,  \label{0c}
\end{equation}
where $\mid 0\rangle $ is the state-vector of the vacuum and $%
x^{1},x^{2},x^{3},x^{4}$ the coordinates of a space-time point, which I
shall label collectively $x^{\eta }$ in the following. However the existence
of quantum fluctuations implies that the vacuum expectation of the product
of the components at two space-time points may not be zero, that is 
\begin{equation}
\left\langle 0\left| \widehat{T}_{\mu }^{\nu }\left( x^{\eta }\right) 
\widehat{T}_{\sigma }^{\lambda }\left( x^{\zeta }\right) \right|
0\right\rangle \neq 0.  \label{0d}
\end{equation}

As said above I shall ignore the effects of matter and radiation, that is I
will consider an empty universe where the quantum vacuum fulfils eqs.$\left( 
\ref{0c}\right) $ and $\left( \ref{0d}\right) .$ The effect of the quantum
vacuum on the curvature of spacetime should be calculated within the
framework of quantized gravity. This means assuming that the vacuum is
characterized by a metric tensor operator $\widehat{g}_{\mu \nu }\left(
x^{\eta }\right) $ which is related to the energy-momentum tensor operator $%
\widehat{T}_{\mu }^{\nu }\left( x^{\eta }\right) $ by some equations not yet
known. An obvious constraint on these equations is that they will agree with
Einstein\'{}s equations of general relativity in the classical limit, that
is 
\begin{equation}
G_{\mu }^{\nu }\equiv R_{\mu }^{\nu }-\frac{1}{2}g_{\mu }^{\nu }R=8\pi
GT_{\mu }^{\nu },\;R\equiv R_{\lambda }^{\lambda },  \label{E}
\end{equation}
where $R_{\mu }^{\nu }$ is Ricci\'{}s tensor. But before going to quantized
gravity let us recall a few results of the classical theory which will be
useful in what follows.

In classical general relativity the Ricci tensor, and hence the Einstein
tensor, $G_{\mu }^{\nu },$ is related to the metric coefficients, $g_{\sigma
\lambda },$ and their derivatives by well known equations of Riemann
geometry, which we might write 
\begin{equation}
G_{\mu }^{\nu }=G_{\mu }^{\nu }\left[ g_{\sigma \lambda }\right] ,  \label{F}
\end{equation}
meaning that $G_{\mu }^{\nu }$ is a functional of $g_{\sigma \lambda },$ the
functional involving first and second derivatives with respect to the
coordinates, $x^{\eta },$ combined with algebraic operations. In principle
the functional may be inverted so that the metric might be obtained as a
functional of $G_{\mu }^{\nu }$, and hence, using eq.$\left( \ref{E}\right)
, $ as a functional of $T_{\mu }^{\nu },$ which we may write 
\begin{equation}
g_{\sigma \lambda }=g_{\sigma \lambda }\left[ GT_{\mu }^{\nu }\right] .
\label{F1}
\end{equation}
It is plausible to assume that, at least in weak gravitational fields, the
functional eq.$\left( \ref{F1}\right) $ may be approximated by a polynomial
in powers of Newton\'{}s constant, $G$, the zeroth order term giving
Minkowski\'{}s metric.

As an illustrative example let us consider a space-time with spherical
symmetry. We may use standard (or curvature) coordinates with metric 
\begin{equation}
ds^{2}=A\left( r,t\right) dr^{2}+r^{2}d\theta ^{2}+r^{2}\sin ^{2}\theta
d\phi ^{2}-B\left( r,t\right) dt^{2}.  \label{dsc}
\end{equation}
Then one of the relations eqs.$\left( \ref{F}\right) $ is\cite{Synge} 
\begin{equation}
8\pi G\rho \left( r,t\right) \equiv G_{4}^{4}=r^{-2}\left( 1-\frac{d}{dr}%
\left( \frac{r}{A\left( r,t\right) }\right) \right) .  \label{43}
\end{equation}
The equations for the remaining $G_{\mu }^{\nu }$ are more involved and will
not be written here. Eq.$\left( \ref{43}\right) $ may be inverted allowing
to get the metric coefficient $A$ in terms of the density $\rho $, that is 
\begin{equation}
A=\left( 1-\frac{2Gm}{r}\right) ^{-1},\;m\left( r,t\right) \equiv
\int_{0}^{r}4\pi x^{2}\rho \left( x,t\right) dx.  \label{A1}
\end{equation}
This expression may be expanded in powers of $G$ and we get 
\begin{equation}
A=1+\frac{2Gm}{r}+\frac{4G^{2}m^{2}}{r^{2}}+O(G^{3}).  \label{A2}
\end{equation}

Let us now pass to the treatment of the large-scale properties of the
universe according quantized gravity. We may expect that a complete quantum
gravity theory, not yet available, would provide relations between the
metric tensor operator, $\widehat{g}_{\sigma \lambda },$ and the
energy-momentum tensor operator, $\widehat{T}_{\mu }^{\nu },$ which, by
analogy with eq.$\left( \ref{F1}\right) ,$ might be written 
\begin{equation}
\widehat{g}_{\sigma \lambda }=\widehat{g}_{\sigma \lambda }\left[ G\widehat{T%
}_{\mu }^{\nu }\right] .  \label{G1}
\end{equation}
Now I assume that the functional $\widehat{g}_{\sigma \lambda }\left[ G%
\widehat{T}_{\mu }^{\nu }\right] $ may be approximated by a polynomial in
powers of $G$. If we calculate the vacuum expectation value of that
polynomial, the term of zeroth order will give Minkowski\'{}s metric. The
term of order $G$ will derive from the presence of matter, any possible
contribution of dark energy vanishing in view of our assumption eq.$\left( 
\ref{0c}\right) $. Thus we expect that \textit{the relevant term of the
vacuum expectation of the metric associated to the vacuum fluctuations will
be of order }$G^{2}.$ This is essentially the result that I wanted to prove.
In the rest of the letter I provide an illustrative calculation involving
some plausible assumptions. The first of these will be that the global
properties of space-time may be obtained from the vacuum expectation value
of the metric tensor operator, that expectation being treated as if it was a 
\textit{classical} metric tensor. That is I will assume that the following
(classical, i.e. c-number) metric tensor 
\begin{equation}
g_{\mu \nu }\left( x^{\eta }\right) =\left\langle 0\left| \widehat{g}_{\mu
\nu }\left( x^{\eta }\right) \right| 0\right\rangle ,  \label{0f}
\end{equation}
determines the global properties of space-time. Obviously the quantum
fluctuations of the metric cannot be derived from $g_{\mu \nu }$. In
particular 
\[
\left\langle 0\left| \widehat{g}_{\mu \nu }\left( x^{\eta }\right) \widehat{g%
}_{\lambda \sigma }\left( x^{\zeta }\right) \right| 0\right\rangle \neq
g_{\mu \nu }\left( x^{\eta }\right) g_{\lambda \sigma }\left( x^{\zeta
}\right) . 
\]
(Furthermore the left hand side of that expression may possess an imaginary
part because the operators involved may not commute.)

I want to reproduce the large scale properties of a de Sitter universe whose
Robertson-Walker metric, with flat spatial slices, may be written 
\[
ds^{2}=a\left( t\right) ^{2}\left( dr^{2}+r^{2}d\theta ^{2}+r^{2}\sin
^{2}\theta d\phi ^{2}\right) -dt^{2}, 
\]
with 
\[
a(t)\varpropto \exp \left( Ht\right) ,\;H\equiv \sqrt{8\pi G\rho _{DE}/3}. 
\]
For our purposes it is more convenient to use standard (curvature)
coordinates so that the metric becomes of the form of eq.$\left( \ref{dsc}%
\right) $\cite{Rich} with coefficients 
\begin{equation}
A=1+\left( 8\pi /3\right) Gr^{2}\rho _{DE}+O\left( r^{4}\right)
,\;B=1-\left( 8\pi /3\right) Gr^{2}\rho _{DE}+O\left( r^{4}\right) .
\label{AB}
\end{equation}
Our aim will be to show that the vacuum expectation of the metric operator,
deriving from the gravity of the vacuum fluctuations, has the form eqs.$%
\left( \ref{dsc}\right) $ and $\left( \ref{AB}\right) .$ Actually the
isotropy of space plus the freedom to choose the radial coordinate implies
that the vacuum expectations of the metric tensor operator, eq.$\left( \ref
{0f}\right) ,$ may be written 
\begin{eqnarray}
\left\langle 0\left| \widehat{g}_{rr}\left( x^{\eta }\right) \right|
0\right\rangle &=&A\left( r,t\right) ,\;\left\langle 0\left| \widehat{g}%
_{\theta \theta }\left( x^{\eta }\right) \right| 0\right\rangle
=r^{2},\;\left\langle 0\left| \widehat{g}_{\phi \phi }\left( x^{\eta
}\right) \right| 0\right\rangle =r^{2}\sin ^{2}\theta ,  \nonumber \\
\left\langle 0\left| \widehat{g}_{tt}\left( x^{\eta }\right) \right|
0\right\rangle &=&B\left( r,t\right) ,\;\left\langle 0\left| \widehat{g}%
_{\mu \upsilon }\left( x^{\eta }\right) \right| 0\right\rangle =0\text{ for }%
\mu \neq \nu .  \label{AB2}
\end{eqnarray}
What remains is to show that the functions \textit{A} and \textit{B} fulfil
eqs.$\left( \ref{AB}\right) .$

In classical gravity the Einstein equations associated to the metric eq.$%
\left( \ref{dsc}\right) $ are rather simple (see e. g. eq.$\left( \ref{43}%
\right) )$ and, furthermore, it is possible to get explicitly the metric
coefficients in terms of the density, $\rho \equiv T_{0}^{0},$ and the
pressure, $p\equiv -T_{1}^{1}=-T_{2}^{2}=-T_{3}^{3}$ (the latter equality
involves assuming local isotropy, that is equality of radial and transverse
pressure). Going to the corresponding equations for quantum operators is not
trivial. In particular, although the vacuum expectation of the tensor
operator $\widehat{g}_{\mu \nu }$ is diagonal, see eq.$\left( \ref{AB2}%
\right) ,$ the tensor itself is not diagonal. Thus the quantum relation eq.$%
\left( \ref{G1}\right) $ is necessarily more involved than the classical
counterpart, eq.$\left( \ref{F1}\right) .$ In the absence of a quantum
gravity theory providing the correct expression, we may plausibly assume
that a relation similar to eq.$\left( \ref{A2}\right) $ is valid in the
quantum case provided operators $\widehat{\rho },$ $\widehat{A},$ $\widehat{m%
}$ are substituted for the quantities $\rho ,$ $A,$ $m,$ respectively. That
is I will assume

\begin{equation}
\widehat{A}\left( r,t\right) \simeq 1+G\frac{2\widehat{m}\left( r,t\right) }{%
r}+G^{2}\frac{4\widehat{m}\left( r,t\right) ^{2}}{r^{2}}+O\left(
G^{3}\right) ,  \label{30}
\end{equation}
where 
\[
\widehat{m}\left( r,t\right) =\int_{\left| \mathbf{x}\right| <r}\widehat{%
\rho }\left( \mathbf{x},t\right) d^{3}\mathbf{x.} 
\]
The choice eq.$\left( \ref{30}\right) $ is by no means obvious, but it is
simple, presents no problem of commutativity of operators and has an
appropriate classical limit. I hope that it is, at least, a good enough
approximation. Neglecting terms $O\left( G^{3}\right) $ it leads to the
vacuum expectation value 
\begin{eqnarray}
\left\langle 0\left| \widehat{A}\left( r,t\right) \right| 0\right\rangle
&\simeq &\left\langle 0\left| 1+G\frac{2\widehat{m}\left( r,t\right) }{r}%
+G^{2}\frac{4\widehat{m}\left( r,t\right) ^{2}}{r^{2}}\right| 0\right\rangle
\nonumber \\
&=&1+\frac{4G^{2}}{r^{2}}\left\langle 0\left| \widehat{m}\left( r,t\right)
^{2}\right| 0\right\rangle ,  \label{30a}
\end{eqnarray}
where the latter equality follows from eq.$\left( \ref{0c}\right) .$

The vacuum expectation may be rewritten as 
\begin{eqnarray}
\left\langle 0\left| \widehat{m}\left( r,t\right) ^{2}\right| 0\right\rangle
&=&\left\langle 0\left| \int_{\left| \mathbf{x}\right| <r}\widehat{\rho }%
\left( \mathbf{x},t\right) d^{3}\mathbf{x}\int_{\left| \mathbf{y}\right| <r}%
\widehat{\rho }\left( \mathbf{y},t\right) d^{3}\mathbf{y}\right|
0\right\rangle  \nonumber \\
&=&\int_{\left| \mathbf{x}\right| <r}d^{3}\mathbf{x}\int_{\left| \mathbf{y}%
\right| <r}d^{3}\mathbf{y}\left\langle 0\left| \widehat{\rho }\left( \mathbf{%
x},t\right) \widehat{\rho }\left( \mathbf{y},t\right) \right| 0\right\rangle
.  \label{30b}
\end{eqnarray}
The correlation of the vacuum density fluctuations should be a function of
the relative distance (at equal times) that is 
\[
\left\langle 0\left| \widehat{\rho }\left( \mathbf{x},t\right) \widehat{\rho 
}\left( \mathbf{y},t\right) \right| 0\right\rangle =f\left( \left| \mathbf{x}%
-\mathbf{y}\right| \right) . 
\]
We do not know the function $f$, but it is possible to reproduce the first
eq.$\left( \ref{AB}\right) $ with the choice 
\[
f\left( z\right) =Cz^{-2}h\left( z\right) ,\;h\left( z\right) =1+O\left(
z^{2}\right) . 
\]
We should assume that the funcion $h(z)$ dereases rapidly to zero when $%
z\rightarrow \infty $ but its exact form is not needed for our purposes. The
integrals eq.$\left( \ref{30b}\right) $ are straightforward for the leading
term at short distances. In fact, writing 
\[
\left| \mathbf{x}-\mathbf{y}\right| ^{2}=x^{2}+y^{2}-2xy\cos \theta , 
\]
the angular integral leads to 
\[
\left\langle 0\left| \widehat{m}\left( r,t\right) ^{2}\right| 0\right\rangle
=16\pi ^{2}C\int_{0}^{r}xdx\int_{0}^{r}ydy\log \frac{x+y}{\left| x-y\right| }%
+O\left( r^{6}\right) . 
\]
Hence we get 
\[
\left\langle 0\left| \widehat{A}\left( r,t\right) \right| 0\right\rangle
=1+16\pi ^{2}G^{2}Cr^{2}+O\left( r^{4}\right) , 
\]
which agrees with the first eq.$\left( \ref{AB}\right) $ if we choose the
constant $C$ so that 
\[
\rho _{DE}=6\pi GC. 
\]
Thus I have shown that plausible assumptions may lead to the dark energy
being the product of Newton constant $G$ times a constant $C$ fixing the
scale of the vacuum density two-point function at short distances.

A calculation of the metric coefficient $B(r,t)$ would be more involved and
also it would require additional hypotheses. For these reasons it will not
be made here. In any case the following conclusions will be probably true in
a rigorous treatment (to be made when quantum gravity theory is available):

1. Quantum vacuum fluctuations give rise to a vacuum expectation of the
metric tensor which departs from the Minkoswski expression. This means that
we should expect that the quantum vacuum fluctuations produce some curvature
of space-time even if the vacuum expectation value of the quantum
energy-momentum tensor vanishes.

2. The curvature of space-time (i. e. the vacuum expectation of the metric
tensor quantum operator) mimics the one produced by some classical
energy-momentum tensor with density and pressure fulfilling $p=-\rho .$

3. The said classical energy-momentum tensor may be written as the product
of Newton\'{}s constant, $G$, times some expression involving the quantum
vacuum fluctuations of the energy-momentum.

These conclusions \textit{suggest }that the dark energy (or mass) density, $%
\rho _{DE}$, and pressure, $p_{DE}$, are fictitious but the curvature of
space-time is real and it is the same that would be produced by a classical
mass density and a pressure as in eq.$\left( \ref{1}\right) .$ If this is
the case the value of that mass density, $\rho _{DE}$, should be obtained as
a product of Newton\'{}s constant, $G$, times some factor, $C$, which
depends on the properties of the vacuum quantum fields, likely those of the
standard model of elementary particles. Thus we might estimate the order of
magnitude of the parameter $C$ by means of a dimensionally correct
combination of the Planck constant, $
\rlap{\protect\rule[1.1ex]{.325em}{.1ex}}h%
$, the speed of light, $c$, and a typical mass of elementary particles, $m$.
Consequently, in order that $\rho _{DE}$ has dimensions of energy density,
we shall assume that its value is given by eq.$\left( \ref{40}\right) .$

It is also possible that the quantum vacuum fluctuations produce only a part
of the dark energy, another contribution coming from other mechanisms. In
fact many mechanisms have been proposed\cite{Copeland}, but most of them
have difficulties in explaining the actual value of the dark energy density.


\begin{thebibliography}{9}
\bibitem{Sahni}  Varun Sahni, Dark matter and Dark energy, \textit{Lect.
Notes Phys.} \textbf{653}, 141-180 (2004). Archive astro-ph/0403324v3.

\bibitem{WMAP}  G. Hinshaw et al., \textit{Ap. J. Suppl.\ Series}, \textbf{%
180 }(2), 225 (2009); Archive astro-ph/0803.0732.

\bibitem{W}  S. Weinberg, \textit{Rev. Mod. Phys.} \textbf{61}, 1 (1989).

\bibitem{Z}  Ya. B. Zel\'{}dovich, \textit{JEPT Lett}. \textbf{6}, 316
(1967); \textit{Sov. Phys. Usp.} \textbf{11}, 381 (1968).

\bibitem{pad}  T. Padmanabhan, \textit{Class. Quant. Grav.} \textbf{22},
L107 (2005); R. Katti, J. Samuel and S. Sinha, \textit{Class. Quant. Grav.} 
\textbf{26}, 135018 (2009).

\bibitem{Z2}  Ya. B. Zel\'{}dovich, \textit{Sov. Phys. Usp.} \textbf{24},
216 (1981).

\bibitem{Synge}  J. L. Synge, \textit{Relativity. The general theory},
North-Holland Pub. Co., Amsterdam, 1965. Page 270.

\bibitem{Rich}  J. Rich, \textit{Fundamentals of Cosmology},
Springer-Verlag, Berlin, 2001. Page 126.

\bibitem{Copeland}  E. J. Copeland, M. Sami and S. Tsujikawa, \textit{Int.
J. Mod. Phys. D} \textbf{15}, 1753 (2006); Archive hep-th/0603057 (2006).
\end{thebibliography}
\end{document}